\documentclass[a4paper,11pt]{article}
\pdfoutput=1 

\usepackage{jheppub} 

\usepackage{natbib}
\usepackage{tabulary}
\usepackage{amsmath}    
\usepackage{graphicx}   
\usepackage{verbatim}   
\usepackage{color}      
\usepackage{subfigure}  
\usepackage{hyperref}   
\title{Underground physics without underground labs: large detectors in solution-mined salt caverns}
\author{Benjamin Monreal}
\affiliation{University of California, Santa Barbara}
\emailAdd{bmonreal@physics.ucsb.edu}

\abstract{A number of current physics topics, including long-baseline neutrino physics, proton decay searches, and supernova neutrino searches, hope to someday construct huge (50 kiloton to megaton) particle detectors in shielded, underground sites.    With today's practices, this requires the costly excavation and stabilization of large rooms in mines.    In this paper, we propose utilizing the caverns created by the solution mining of salt.   The challenge is that such caverns must be filled with pressurized fluid and do not admit human access.    We sketch some possible methods of installing familiar detector technologies in a salt cavern under these constraints.   Some of the detectors discussed are also suitable for deep-sea experiments, discussed briefly.    These sketches appear challenging but feasible, and appear to force few major compromises on detector capabilities.   This scheme offers avenues for enormous cost savings on future detector megaprojects.}

\begin{document}

\maketitle
\flushbottom

Our knowledge of particle physics would be advanced in several directions if we could build densely-instrumented, deep-underground detectors on the 50 kT-MT scale.   Topics include\footnote{High-energy cosmic-ray detectors like IceCube, which are much larger than anything we consider here, will be excluded them from this discussion.} long-baseline neutrino physics, atmospheric neutrinos, supernova neutrinos, proton decay, reactor neutrinos, and geoneutrinos.  Such detectors prefer to be installed underground for cosmic-ray shielding, but the mining costs for 50 kT--MT caverns are prohibitively large.   In addition, there are substantial costs associated with building and operating basic mine infrastructure (hoists, dewatering, safety, etc..).   We currently have facilities, plans, and in some cases cost estimates for an upcoming generation of underground laboratories and detectors~\cite{deGouvea:2013onf} with total project costs approaching or exceeding \$1 billion.   Thinking ahead to the subsequent generation, is there any way to get a large detector underground without the mining costs?   

\begin{figure}
\begin{center}\includegraphics[width=0.6\linewidth]{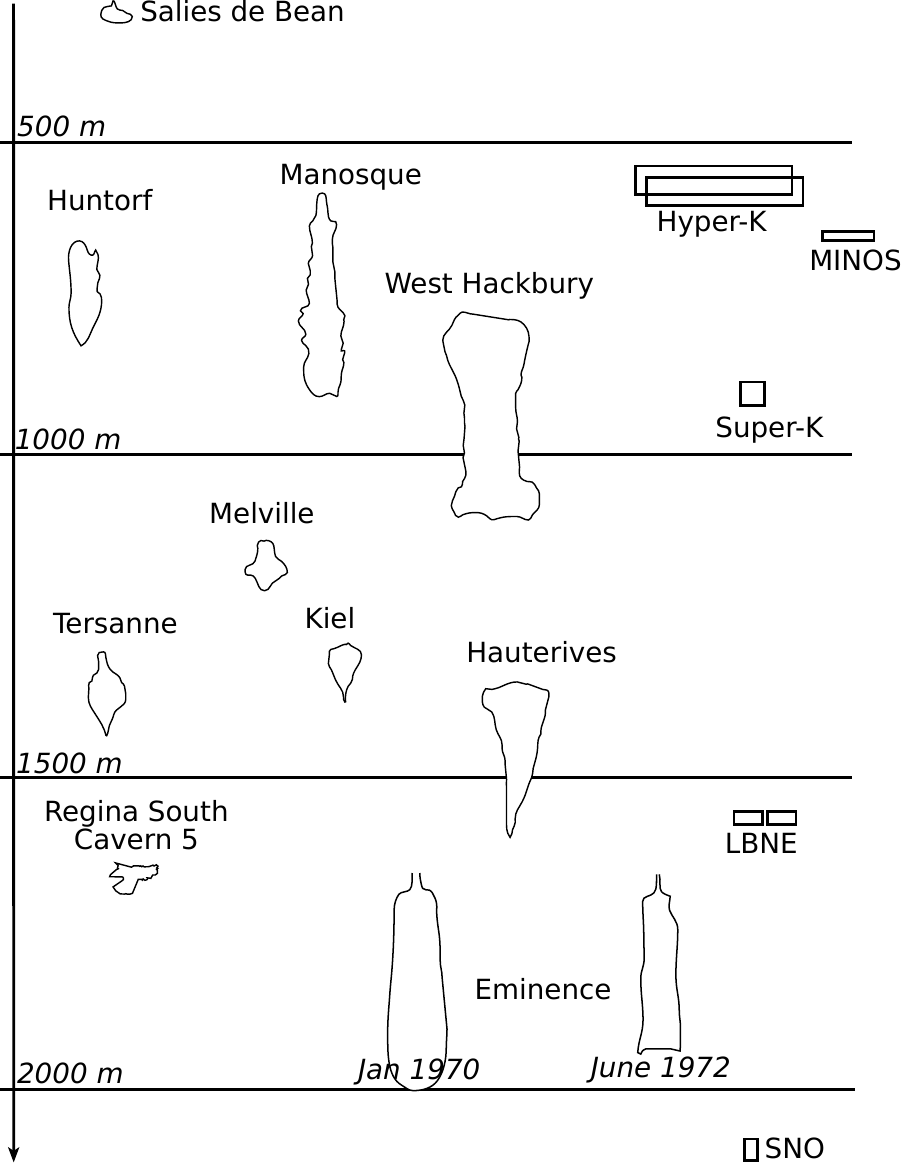}\end{center}
\caption{Examples of existing salt caverns and selected underground lab excavations (real and proposed) for scale. Two drawings of the Eminence cavern show its rapid shrinking due to salt creep when underpressurized.  Cavern data from~\cite{Warren2006}).}\label{ex_cav}
\end{figure}

The first objective of this paper is to point out the existence of an extremely cheap method of creating underground caverns.  This is \emph{solution mining}.   Solution mining is a low-cost way to recover valuable salts (usually halite, but also potash and other salts) from deep salt formations, leaving behind an enormous cavern in the salt.   The mining is carried out by drilling and casing a shaft into a salt formation, most commonly commonly 500--2000~m deep.   A concentric pipe is lowered into the shaft and fresh water is delivered down the pipe; the fresh water dissolves salt from the cavern walls, becoming a nearly-saturated brine which returns to the surface via the annular space between the pipe and the well casing.   The volume and span of the resulting caverns can be stupendous; roof spans as large as 366 m and volumes up to $10^6$ m$^3$ are reported.   (See figure~\ref{ex_cav} for some examples.)   Caverns, some left over from former salt-extraction and some purpose-built, are commonly used as storage tanks for petroleum products or natural gas, or more recently for compressed-air energy storage.   The cost of such mining is discussed in section~\ref{costs} and is a small fraction of the cost of conventional excavation.  

But these are just caverns, not mines.   The caverns are constantly under elevated pressure and temperature, and lack the geotechnical engineering (roof bolting, shotcreting, ventilation, etc.) that make mines safe for human entry.   This drives two lines of inquiry.  First, are these caverns safe, stable, and controllable enough to serve as a site for a costly and high-priority detector project?  Second, can we actually install a large detector in a pressurized, inaccessible space?    In this article we argue that both answers are ``yes'', though of course much research and engineering is needed.  

\section{Motivation: Physics and project-planning considerations}

The main motivation behind the salt-cavern approach would be the potential for very low construction and operation costs, compared to similar  conventionally excavated mines and rooms.    There may also be site-selection benefits.   Finally, salt caverns may be suitable for housing huge high-pressure gas time projection chambers (TPCs) which are impossible to build in conventional mines.

\subsection{Facility cost savings}\label{costs}

The excavation and operation costs associated with mines are very large, even by the standards of modern physics megaprojects.    Some high-cost items worthy of comment here include shaft sinking, waste rock hoisting and disposal, cavern excavation, and cavern support.    Large cavern  costs in the US can be in the range \$500-\$1000/m$^3$; shaft sinking costs, for the large shafts (7--8 m) used for worker access, mucking, and ventilation, can be \$30k/m.      Some selected project cost estimates are summarized in Table \ref{tabcosts}.    Operating a mine is also expensive: the Sanford Lab has an operating budget of \$20M/y.

\begin{table*}
  \begin{tabular}{|p{0.15\textwidth}|p{0.35\textwidth}|p{0.25\textwidth}|p{0.25\textwidth}|}
\hline
Project & Detector & Facility cost estimate & Total project cost \\
\hline
LBNE-WC & 200 kT water at 1530 m & 960 M\$ & 1400 M\$~\cite{energy.gov}\\
LNBE-LAr & 34 kT argon at 250 m&  640 M\$ & 1330 M\$~\cite{energy.gov}\\
Hyper-K & 1 MT water at 650 m & 400 M\$ & 800 M\$~\cite{hyperk_cost}\\
LENA & 50 kT scint at 1450 m & 100 M\$ & 530 M\$ ~\cite{lena_cost}\\
\hline
\end{tabular}
\caption{Recent cost estimates for the largest detector proposals.}\label{tabcosts}
\end{table*}

By contrast, salt caverns are created for something closer to \emph{a few dollars} per cubic meter.   As part of the National Strategic Petroleum Reserve (SPR), a 14-cavern 13~Mm$^3$ hydrocarbon-storage facility was built in 1992 at Big Hill, TX, for \$270M (including aboveground gas/pipeline infrastructure); this amounts to \$21/m$^3$ overall, of which \$2/m$^3$ is quoted as the marginal cost of cavern expansion~\cite{Warren2006}.    The entire SPR, consisting of four sites, 63 caverns totalling 86~Mm$^3$, and all the associated petroleum infrastructure, has a \$130M/y M\&O budget.~\cite{spr_contract}.   The small shaft-drilling operations we will discuss are inexpensive, too, with costs closer to \$5k/m than to \$30k/m.

With large possible savings in shaft, operations, and especially cavern-excavation budgets, solution-mining may be a path to capital cost reductions on the order of 50\%, or hundreds of millions of dollars, in the cost of building next-generation underground experiments.   

\subsection{Detector site flexibility}
One remarkable feature of salt formations is their ubiquity; thick salt beds and salt domes are extremely common worldwide, giving project planners a fairly wide freedom of site selection.   For example, the planned LBNE neutrino beamline will be (by accident) aimed roughly towards the Jurassic-age Pine Salt formation, which extends into northwestern South Dakota in a formation 60-100m thick and 300-600m below the surface~\cite{zieglar1956pre}.  The southern edge of the Pine Salt approaches within $\sim$40 km of Lead, SD, allowing the possibility of a salt lab ~2$^\circ$ (40 mrad) off-axis from a beam aimed at the Sanford Lab from Fermilab; this opens the possibility that a salt cavern experiment could capitalize on this large beamline investment.    Large fields of caverns already exist at at useful long-baseline distances from several accelerators, including in Texas, Louisiana, Michigan, northern Germany, southern France, and near existing underground labs at WIPP, Boulby and CanFranc.    

Salt formations exist in the Southern Hemisphere in Australia and Argentina, at latitudes which presently have no underground-lab infrastructure; this could be useful for dark-matter searches because southern-hemisphere replication may be a cross-check for claimed annual-modulation observations~\cite{DAMA,COGENT}.

Geoneutrino experiments are ``small'' by the standards of this discussion, but may need an unusual level of site flexibility.   Geoneutrino detection can be usefully accomplished in 10--20~kT scintillator experiments.  The expense is not onerous to excavate a \emph{single} mine cavern for such an experiment (and indeed several already exist) but geoneutrino data is most useful if we have a wide survey of different parts of the Earth's crust and mantle, particularly including sites far from nuclear reactors.    It is possible that an appropriately-designed geoneutrino detector (particularly an extra-small design that fits down existing, unmodified wells) could affordably ``tour'' the Earth's crust inexpensively by visiting a series of salt caverns.    In section~\ref{ocean} we briefly note opportunities for deploying salt-cavern-developed technologies on the seafloor, where the more interesting mantle geoneutrino signal is accessible.

\subsection{Detector technology options and cost savings}
In some cases, the constraints of this cheap cavern will force us to spend more on an otherwise-conventional detector---we will be forced to add pressure enclosures to PMTs, for example---in order to do the \emph{same} physics as we'd do in an unpressurized lab.   But this is not always true.   For example, a large fraction of the cost (30--50\%) of a large liquid-argon TPC is in cryostats and cooling.  In section~\ref{lar} we sketch the possibility of a high-pressure, high-density gas TPC.   Lacking a cryostat and cryogenics, the detector itself may be cheaper per unit mass than is possible on the surface, and the somewhat-lower target density may actually improve the detector performance.   A large component of the budget of NOvA is devoted to the rigid, load-bearing PVC structure; the salt cavern version described in section~\ref{ls} is buoyancy-supported, allowing cost savings on this budget item which may offset complexities added elsewhere.

\section{Constraints: Mining and geotechnical considerations}

There is a large literature on the geology, engineering, economics, and risks of solution mining. Here (relying on~\cite{Warren2006} unless otherwise noted) we review a few of the points that appear most relevant to nuclear and particle physics.

\subsection{Cavern shapes}

The shape of the cavern is determined by details of the flow of fresh water over the salt.

\begin{enumerate}
\item The cavern engineer has control over where fresh water is injected, and where brine is withdrawn, along the vertical span of the cavern.   Injection at the bottom of the cavern tends to enlarge the bottom, injection at the top tends to enlarge the top.
\item The engineer can prevent water from contacting the cavern roof by inserting a buffer layer of oil or gas that floats over the water.   Injection of fresh water at the top of the cavern, underneath such a buffer, tends to dissolve the walls (but not the ceiling) near the top of the cavern.
\end{enumerate}

The latter process in particular allows the construction of a cavern of almost any desired cylindrical profile.    This process is most precise in large, uniform bodies of halite, typically salt domes.   In bedded salt formations, one might instead find a series of thin strata (not necessarily horizontal) of varying composition and solubility, and here the cavern shape control will be much less precise.   Most caverns begin and end within the salt body, rather than attempting to remove salt right up to the insoluble caprock; this aids cavern stability, particularly because in many cases the caprock is a mechanically-weak anhydrite.    All salt formations contain some insoluble inclusions (sand, pebbles, anhydrite) which accumulates as a bed, along with some undissolved salt, on the floor of the cavern.  The unconsolidated nature of the floor should be taken into account when designing anchoring schemes later.

\subsection{Cavern stability and pressure}
All underground rock is under stress due to the lithostatic pressure $P_L$ exerted by  the weight of rock overburden.  As in all excavations, we have removed some of the load-bearing rock, which transfers stress to the cavern walls.   The maximum safe size and shape of caverns is determined by the compressive strength of rock under this stress.   In contrast to mined rooms, in solution-mined caverns the rock is replaced with a pressurized fluid which helps balance $P_L$.   This pressure is key to understanding the long-term maintainence of deeper caverns.   Salt caverns have a problem other than collapse: salt undergoes creep deformation in response to stress, particularly at high temperatures and pressures, so even a cavern which is stable against fracture will gradually close up unless the internal pressure balances $P_L$.    A easily-maintained, fairly-constant pressure can be obtained by simply filling the cavern and the well with saturated brine, resulting in ``halmostatic'' pressure $P_H = \rho g h$ in the cavern, where $\rho$ is the brine density, $g$ is the acceleration due to gravity, and $h$ is the depth.    Since brine is less dense than rock, $P_H$ only incompletely balances $P_L$ and will slow, but not stop, the creep.    The cavern closure timescale under $P_H$ is site-dependent, but values seen in the literature include 10$^5$ y in a cavern at 250m depth~\cite{brouard2013},  10$^4$y (with ``large variations'') at 1000m~\cite{berest2010}, but 10--100y at 2000--1400m~\cite{coates}, all but the shortest of which appear compatible with a detector site.   Higher pressures can be maintained by using a fluid column denser than brine (like a barite suspension ``mud'') in the shaft, or by sealing the shaft at the wellhead and applying additional mechanical pressure on the whole shaft/cavity fluid.  (An abandoned, sealed, leaktight cavern will increase in pressure on its own, as the salt creeps and compresses the fluid.)  In this paper, we assume---although this needs to be checked on a case-by-case basis---that the cavern is stable enough to be left at $P_H$ during detector construction, even though it prefers a higher pressure during long-term operations.  If this is not true, the mud and/or the pressurization hardware will be a notable complexity and expense which might be unavoidable for work at greater depths.  

Caverns may also collapse via failures of the roof.    Large-scale roof integrity has been heavily studied by the mining and storage industry, since a sequence of roof collapses (``chimneying'') can lead to a surface sinkhole.    Roof integrity considerations favor narrower caverns with smaller roof spans, but spans of 80-100m appear to be fairly routinely possible and stable for indefinite periods of time.   Numerous available cavity-collapse case studies appear to involve poorly-managed freshwater injection during the creation and use of the cavern, not spontaneous failure of statically brine-filled cavities.

On a smaller scale, rockfall from a cavern roof could damage a detector; indeed, rockfall prevention is one of the particular tasks of creating human-accessible mines.  Falling salt blocks, sometimes large enough to damage a drill string or casing, appear to be common during active dissolution but rarer for a static cavity.  Unfortunately, the details of rate and distribution of smaller rockfalls, being fairly irrelevant to industrial interests, appears to be poorly studied.   We will discuss this further in section~\ref{lining}.

We tentatively conclude that a halmostatic-pressure salt cavern, particularly at at depth shallower than 1000--1500 m, will be suitably long-lived for a large detector project,  and that deeper caverns need should not be rejected out of hand but need more site-specific study.

\subsection{Shafts}
A solution mine is typically created with only a small hole, called a ``well'', of order 10--18 inches diameter.    Since we want to deploy large detector components, we will discuss boring the well into a hole we will call a ``shaft''.   Although we note that large-shaft sinking into a salt cavern does \emph{not} have any precedent that the author is aware of, we see no obvious showstoppers.     If the original solution-mining well is bored out, the shaft will necessarily enter the top dead center of the cavern, but it is equally easy to imagine drilling and boring a new shaft off to one side, or even inclined, if this simplifies detector installation somehow.

The mining industry has extensive experience in boring and casing vertical shafts: for mine access, movement of rock, ventilation, and in some famous cases for rescue.   Drilling a shaft into a salt cavern is an unusual hybrid operation: we have a hole that the rock spoil can fall into (making it resemble raise boring) but we don't have access to the bottom of the hole for installing a boring bit (making it resemble blind boring).    Unlike raise boring we do not have to muck out the spoil at all; it can be abandoned on the floor of the cavern.     Blind boring rigs up to 6m diameter are available\footnote{Readers may be familiar with even larger-diameter vertical shafts, like the 20~m access shafts to LHC experiments.   Such shafts are constructed with continuous human- and machinery-access to the cutting face, which is not possible in a water-filled shaft.}, although in this paper we will focus on the 1.5--2.5~m range.       

\section{Detectors technologies for salt caverns}

\subsection{Overview of engineering constraints}

The constraint imposed by this choice of mine are very different than the constraints of normal underground work:

\begin{itemize} 
\item Our detector must survive under high pressure, at least $P_H$ and possibly higher; we may require construction to occur at this same pressure.   
\item All detector components must fit down a fairly small shaft.
\item The detector must deploy into an irregularly-shaped cavern with minimal geotechnical engineering (support points, cranes, etc.) and notably corrosive surroundings.
\item All detector installation must be done robotically.   
\end{itemize}

If we can meet these constraints, and if they don't impose absurdly-high costs relative to standard detector construction, we might be able to take advantage of a solution-mined cavern.   We note to begin with that some of these operations have precedents in neutrino physics: 

\begin{itemize} 
\item Undersea (Antares, Km3Net) and under-ice (IceCube) detectors operate reliably at high pressure.
\item A number of large detectors have built large articulated objects for overhead deployment via a narrow neck, including source/calibration arms (used in SuperK~\cite{superk_arm}, Borexino~\cite{borexino}, and many others) and cleaning/polishing equipment (used in DEAP~\cite{deap}).
\item The SNO neutral-current detector system was inserted into a spherical detector via a narrow neck, and driven into place using a remotely-operated submersible (ROV)~\cite{amsbaugh2007}, in a procedure very similar to some discussed here.  
\end{itemize}

The discussion that follows is speculative.    We will sketch out several general engineering concepts for outfitting a salt cavern and constructing a large detector inside, subject to the constraints above.   These are only sketches, and we invite interested readers to devise better concepts and/or to discard the unviable aspects of these.    We will start by discussing the infrastructure common to all detectors.  We then examine three well-studied large detectors or proposals (HyperK, NOvA, and LBNE) and several small ones and ask how closely these familiar detectors could be duplicated in a salt cavern.    We do not attempt to recreate the physics-sensitivity nor the detector-properties reasoning that motivated these detector designs to begin with, both topics having been thoroughly surveyed recently~\cite{Hewett:2014qja, deGouvea:2013onf}.   

\subsection{General: Cavern lining and infrastructure}
\subsubsection{Cavern lining}\label{lining}
Most underground detectors are installed in \emph{lined} rock caverns; after the raw cavern is mined, it is reinforced with bolts, lined with concrete and additionally given an impermeable plastic coating, or in some cases a steel tank.  This may sound difficult in an irregular, flooded salt cavern, but some sort of shielding is desirable and in some cases unavoidable: brine is corrosive, natural salts are high in $^{40}$K and radioactive, and water of uncontrolled salt content also has uncontrolled optical properties.   Most detectors will prefer to deploy in a clean and purified fluid, either fresh water or oils; therefore, we need a barrier between the detector fluid and the salt.   Is this possible?     

Consider a 10$^6$ m$^3$ ``carrot-shaped'' cavern 80m diameter, 240m high cylindrical cavern enclosing 1.2 Mt of water.  (This is roughly the size and shape of the \emph{largest} fuel-storage caverns in salt domes.   We discuss this extreme case first; everything will be easier in a smaller cavern.)   A liner for this cavern could be manufactured on the surface in the form of a one-piece elastic balloon.  The balloon is rolled up into a cylinder, lowered down the shaft, and inflated to conform to the cavern walls.    A balloon made of material 1cm thick would, when rolled up, form a 1.8 m diameter bundle and ``inflate'' to line the entire cavern\footnote{The idea of ``inflation'' here presumes that we can access the fluids, and apply different pressures, both inside and outside the balloon.   The fluid outside the balloon could be accessed via an additional pipe within the main shaft or via a separate well.}.   Once such a liner is installed it might be possible to inject grout outside of it to help with ceiling stabilization; if not, the liner itself, with a small pressure deficit maintained on the outside, should be strong enough to trap small rockfall slabs against the roof.    We note that we need not limit the ``balloon'' materials to rubbers and polymers: with very large pressures available, a thin \emph{sheet metal} bladder could be ``inflated'' from a folded or rolled state into a very large tank; the procedure might be something halfway between a work of origami and a work of hydroforming or hydrobulging.   The one-piece liner is far heavier (400~t for a 1~mm steel or 5~mm rubber liner) than anything else we will contemplate installing, and on the high end for hoisting operations generally, so perhaps this load would ``walk'' down the shaft with a pair of hydraulic jacks. )

Alternatives to one-piece liners are even more speculative.  One could deliver a series of thin balloons and inflate one inside the next until the desired material thickness is built up, perhaps also including adhesives in the space between them.   It is possible that large liners could be assembled piecewise underwater, or use an ROV to paint the cavern walls with a thick polymer coating.    

\subsubsection{Mechanical infrastructure}
The lack of general-purpose construction access is an important factor in all of our detector-design considerations.   From the surface, we can perform hoisting or drilling operations only along the centerline of the cavern, directly below the shaft\footnote{We note that a cavern might be penetrated by multiple shafts, which may or may not help.   The one-piece cavern lining described above is only straightforwardly compatible with a single main shaft. }.   Any activities envisioned \emph{off} of the centerline require special attention.   An ROV, swimming in the cavern fluid, can deliver neutrally-buoyant payloads to any desired location in 3D.    ROVs have proven themselves capable, familiar, and cost-effective in marine construction, we generally assume they are the best option for caverns as well, despite some difficulties managing ballast.  There is even a precedent for driving an ROV in a salt cavern~\cite{ballou}.   ROVs do not carry heavy loads, so we need to either restrict ourselves to neutrally-buoyant objects \emph{or} continue searching for a heavy-lifting system that can reach sideways in the cavern.     

A long beam, hanging from two hoist cables, can be levered away from the centerline, although not with very high positioning precision.   If more-complex machinery is permitted in the cavern, a knuckle-boom crane can be delivered down the shaft in one operation; a multistep deployment might be able to install something like a bridge crane.   One can bootstrap one's way from light infrastructure to heavy: for example, an ROV can drag a light guide-cable to one side of the cavern; that cable can be used as a hauldown for a heavier cable; the heavy cable can guide a heavy beam, etc.. Most of the designs discussed do not obviously \emph{need} such equipment.    We note that in many cases strong, massive beams, cranes, etc. can be made neutrally-buoyant for easier ROV work.  

Despite the thin liner, it is possible to have strong wall- and floor-anchors inside the cavern.   A strong point can be preinstalled in the liner balloon, connected to a bundle of anchor chains or cables hanging into the cavern space or perhaps into a ``pocket'' outside the liner.   After the liner is installed, grout can be pumped into the anchor area or into the pocket, resulting in a strong tiedown.   It should be remembered that the cavern floor is unconsolidated sand and rubble, not hard rock, and has no tensile strength. 

\subsubsection{Data, power, fluids, and ROV}
Once the cavern is lined, we can begin installing equipment inside it.   Where does this equipment go and what is needed?   A number of fluid, gas, and electrical conduits will be lowered into the shaft and connected to a manifold inside the cavern.   The ROV, if present, needs some sort of parking place.   A hydraulic manipulator arm (either a stationary one or one of the ROVs') needs  to meet incoming detector components and do work on them---decoupling them from the hoist, connecting cables, etc..    Some such equipment would be left in place during detector operations, and some (with reduced cleanliness requirements) is needed only during construction and can be withdrawn (either to the surface or to an isolated and less-clean fluid subvolume) before operations.   

\begin{figure*} \begin{center}\includegraphics[width=0.8\linewidth]{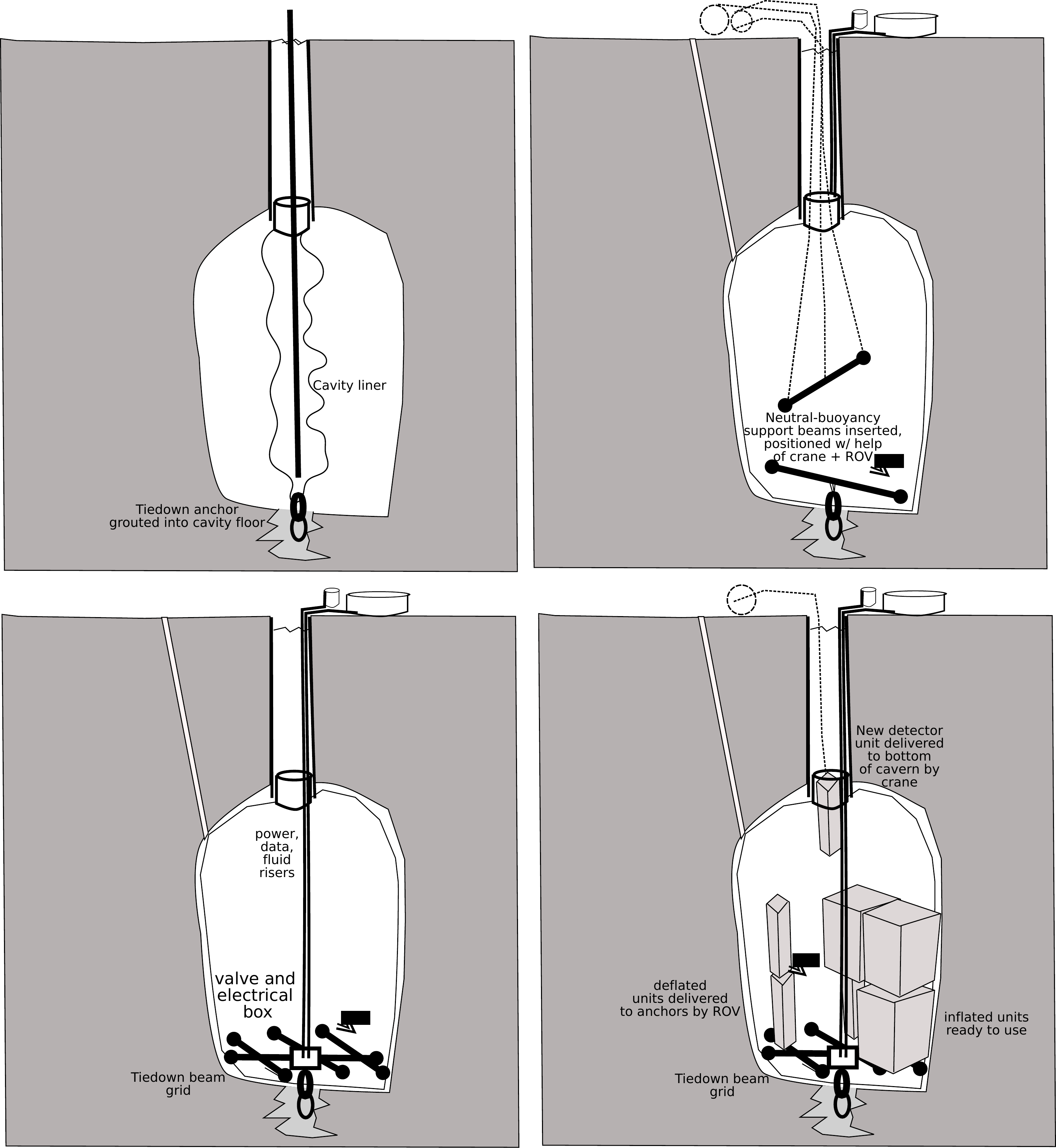}\end{center}
\caption{Cavern preparation and construction sequence for a NOvA-like detector.   Here we show the installation of a strong anchor point and crane-deployment of a tiedown structure, and placement of manifolds and distribution boxes on the cavern floor instead of at the collar.}\label{fig_nova}
\end{figure*}

For experiments that need the cavern to be unobstructed, all of this infrastructure can be parked along the sides of the cylindrical casing at the shaft/cavern interface, which we will refer to as the ``collar'', in the style shown in figure~\ref{HyperK}.    If the experiment doesn't mind cables/pipes passing through the cavern, a base station can be located on the cavern floor as shown in figure~\ref{fig_nova}.    Some liner-installation procedures might allow cables to be pre-routed along the liner wall during its construction on the surface.  
 
\subsection{Water Cerenkov detectors}

There is considerable precedent for installing photomultiplier tubes under high pressure.   Cosmic ray Cerenkov detectors in the Antarctic ice cap (AMANDA, IceCube, later DeepCore, PINGU, etc.)~\cite{ice} and deep underwater (DUMAND, ANTARES, KM3NeT)~\cite{subsea} have placed PMTs and electronics into spherical pressure housings, cabled them together into linear ``strings'', and successfully lowered those strings into high-pressure environments.  The HANOHANO~\cite{hanohano} collaboration has designed a liquid scintillator detector for a one-piece deployment into the ocean, the only non-``string''-based design we are aware of.   The standard procedure is to create a Digital Optical Module (DOM) enclosing the PMT itself, along with a high-voltage supply, amplifier, and digitizer, in a spherical pressure housing.    Spheres are commercially available (Nautilus Marine Services, Teledyne Benthos) up to 17'' diameter; a typical arrangement is a single 10'' PMT in a 13'' sphere, although KM3NeT uses numerous small PMTs per sphere.   We require approximately 400,000 13'' DOMs to match the HyperKamiokande photocathode area.

To turn our extreme 240$\times$80 m cavern (introduced in section~\ref{lining}) into a water Cerenkov detector with HyperK-like optical path lengths, we could divide the cylinder into four pie-shaped quadrants.   The detector is then built out of 1100 strings, each 240~m long and bearing 480 13'' DOMs on 50~cm centers.    With a 1.8~m shaft, as many as 38 such strings could be lowered in a single bundle.    Additional strings would be needed in an external veto region.   In contrast to ice/sea detectors, where the DOMs are meant to have isotropic sensitivity, our PMTs must face ``inward'' with reasonable accuracy, but this can be arranged with an appropriate ladder-like assembly of one or more strings.   Compared to HyperK, it seems likely that the PMT positioning accuracy will be low, but with adequate optical calibration this should not hurt detector performance too badly.    
 
Although ROV installation may sound difficult, we note that the SNO experiment performed a conceptually similar installation during its 3rd phase~\cite{amsbaugh2007}.  Experimenters inserted 40 long, buoyant $^3$He proportional counters into a spherical heavy water vessel via a narrow central neck, then carried them to attachment points using a small ROV.   The 40-string operation lasted 6 months, much of which was due to counter assembly and testing procedures rather than to the challenges of the ROV.

SuperK and similar detectors require optical isolation, but not water isolation, between the inner fiducial volumes (viewed by most PMTs) an outer veto region (between the curtain of PMTs and the cavern walls).   Since we saw in section~\ref{lining} that large one-piece curtains and balloons are challenging, here we suggest a non-one-piece solution.   Consider a 240m long, 50cm diameter cylindrical opaque mylar balloon, initially furled along the length of a string.    After all of the strings are installed, a low-pressure pump on each string inflates the balloon with water, making an opaque tube.  The ensemble of tubes, bunched together behind the DOM strings, forms an optical barrier.  

\begin{figure*}
\begin{center}\includegraphics[width=\linewidth]{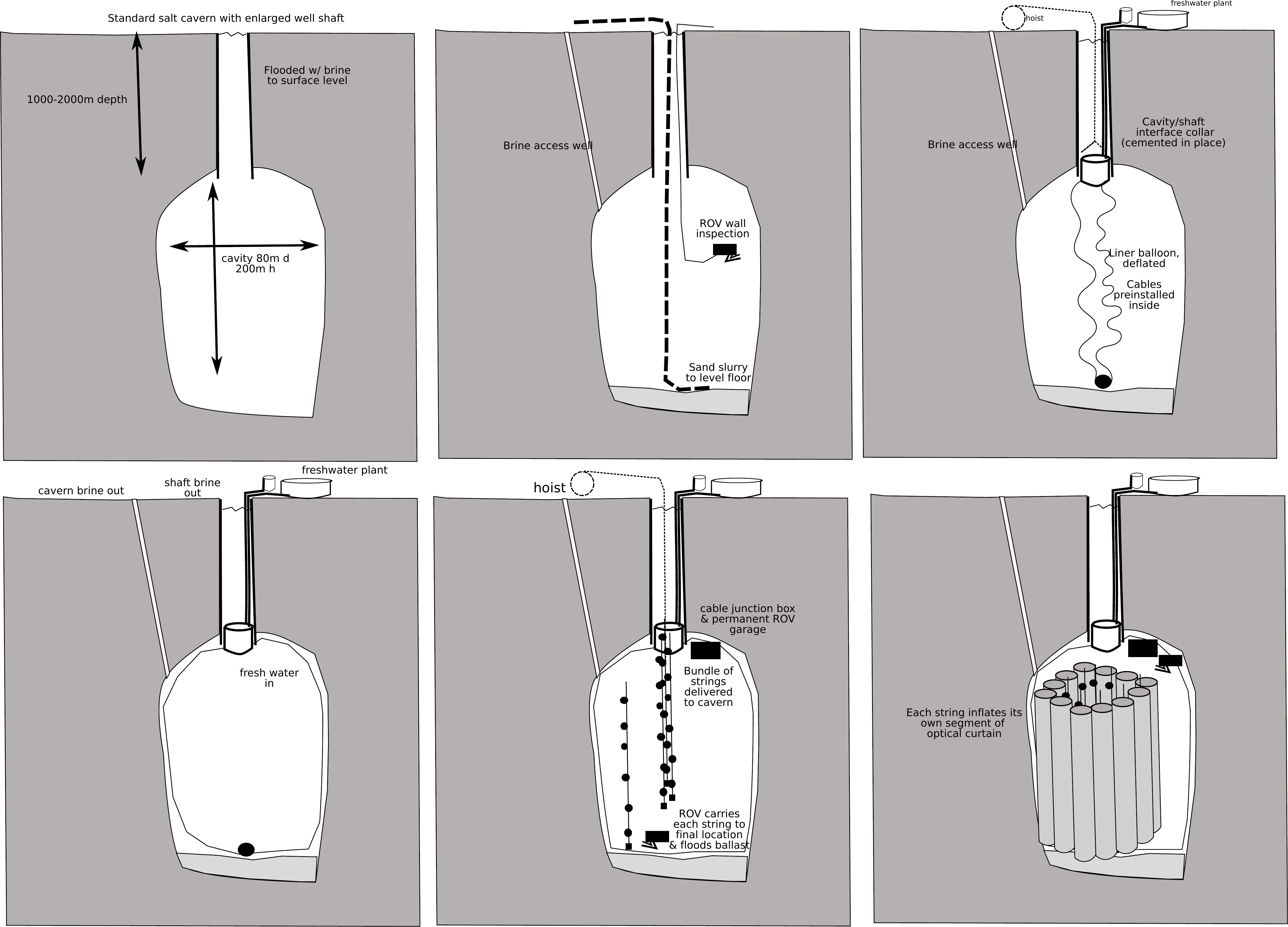}\end{center}
\caption{Cavern preparation and construction sequence for a HyperK-like detector.}\label{HyperK}
\end{figure*}

In figure~\ref{HyperK} we show a cartoon of the deployment sequence for delivering a HyperK-like detector into an initially-bare salt cavern, with a ``simple'' anchoring system whereby strings simply sit on the smoothed cavern floor, held by a ballasted foot.  In later sections we sketch concepts for cemented anchors and rigid tie-downs, which might be necessary here.

This configuration is straightfoward to fill with liquid scintillator; we note recent successes in developing inexpensive water-based scintillators~\cite{Yeh2011} and proposals to deploy them at the 100kT scale~\cite{ASDC}.   It appears somewhat more difficult to deploy, at large scales, the multi-fluid balloon structure used for the lowest-background scintillator experiments~\cite{borexino,kamland} but this arrangement is worth studying.

 \subsection{Liquid scintillator / fiber hodoscopes}\label{ls}

 In contrast to HyperK's large volumes of clear water, the NOvA neutrino detector~\cite{nova} has a finely divided active volume.   NOvA is a hodoscope comprising 500,000 rectangular (4$\times$6~cm) subvolumes filled with liquid scintillator.   Each subvolume is defined by a reflective white PVC tube, within which is a wavelength-shifting fiber with avalanche photodiode (APD) readout on both ends.    This arrangement has interesting advantages for deployment via a tight bottleneck, and for survival at high pressure.   

As built, NOvA's PVC tubes serve as both optical devices and as the detector mechanical structure.   For a submerged and hydrostatically-supported detector, the ``cellular'' structure does not need to be mechanically rigid and indeed may be quite lightweight.   Thin sheets of PVC, Teflon, or Tyvek can be assembled into a collapsible honeycomb structure.     Consider a honeycomb made of 50$\mu$m Tyvek sheets, bonded so as to form into 4$\times$6~cm cells.   Each cell gets a single 1~mm wavelength-shifting fiber (WLS).   This structure compresses to a cross section of less than 10~mm$^2$; with fairly tight compression, a 1.8~m-diameter shaft can accomodate over 100,000 such cells\footnote{NOvA is has alternating planes arranged with a 90$^\circ$ stereo angle.  The tightest folding is only achievable with zero stereo angle, i.e.~with essentially 2-D tracking, which may not be acceptable from a physics standpoint.   The foldability vs. stereo-angle tradeoff needs to be weighed in different physics contexts.}.   
A large block of this honeycomb, enclosed in a watertight bag, serves as the basic detector unit.   The unit is pulled from the collapsed state to the expanded state by filling it with liquid scintillator to a slight overpressure.   We note that this structure might not require a cavern liner, since the detector subvolumes are watertight and could live directly in brine.   If the cavern is unlined, the detector units need some sort of awnings for rockfall protection.

Unlike PMTs, SiPMs and APDs can be directly pressurized.     As long as there is no air space in the packaging, an APD or SiPM board and all the associated electronics can be immersed in degassed mineral oil and subjected directly to high pressures; the electronics need to be enclosed, but only in a (lightweight, cheap) hermitic housing and not a (costly) pressure vessel.    
One attractive option might be to make independently-deployable units comprising 10,000 fibers and 10,000 cells in a $4\times6\times40$ m bag, with one or two electronics boxes (with a 10$\times$10~cm window and SiPM array) at each end, which all together weighs 6--7~t during deployment and inflates to a 800--1000~t target.   Any desired number of such units can be tethered together, vertically or side by side, to reach the desired target mass in the available cavern.    Small stereo angles can be implemented within individual units, or units can be tethered together to form alternating stereo planes on a coarser scale.  The deflated unit may be compatible with a small conventional well bore.

An important question for SiPM or APD-based detector is that of cooling.   SiPMs and APDs have very high dark-count rates at ambient temperatures, and more so at the high ambient temperature (roughly 25$^\circ$/km) underground.    It is possible to cool an entire cavern~\cite{schalge}, taking advantage of the fairly low thermal conductivity of rock.   (At the extreme, cryogenic cooling has been attempted~\cite{haddenhorst} but was found to fracture the rock.)   If we wish to cool the electronics boxes alone, we need insulation; the only thermal insulation capable of high-pressure operation is syntactic foam, with conductivities no better than 0.1~W~m$^{-1}$K$^{-1}$, which appears just adequate for insulating small thermoelectrically cooled boxes.   
 
Relative to a water Cerenkov detector, we face added complexity due to the possibly-large buoyancy of liquid scintillator in water, which demands stronger anchoring systems and pressure-rated balloons holding the subunits.    The solvents which are now most common in large experiments are pseudocumene (PC) and linear alkyl benzene (LAB), both fairly buoyant (890 and 860 kg/m$^3$ respectively) in water; but a mineral-oil fill could provide a good density match.   Phenylxylylethane (PXE) has a density (985 kg/m$^3$) is quite well matched to a freshwater-filled cavern.    On the other hand, the containment balloon can use internal stays (possibly by tension-loading the Tyvek honeycomb itself) to achieve higher pressure ratings than a comparable hollow balloon.


\subsection{Noble gas TPCs}\label{lar}

At first glance, TPCs might appear to be poorly suited for salt-cavern deployment.   Both of the salt cavern's unusual difficulties come into play: unlike PMTs and hodoscopes, drift chambers are \emph{not} so easily divided into small independent functional units; a TPC requires anodes, cathodes, and drift regions of certain sizes in a certain geometry.   The smallest functional unit of the LBNE design is a $7\times2.5\times7.4$~m rectangle, which is a poor shape to fit down a shaft, but nontrivial to reshape since both long dimensions are relevant to the functionality\footnote{The 7~m dimension sets the length of anode wires and therefore the number of electronics channels; the 7.4~m dimension is twice the drift length.}.   Secondly, modern massive TPCs use cryogenic liquid noble gases, but neither membrane cryostats nor dewars would survive under very high external pressures.

However, cryogenics may be a red herring.  In the deeper caverns available to us, at ambient temperature and halmostatic pressure, argon is a dense, supercritical gas.  At 40$^{\circ}$~C and 100 bar, argon is at 160 kg/m$^3$; cooling the whole cavern (mentioned in section~\ref{ls}) to -20$^{\circ}$~C brings that to 450~kg/m$^3$.  In other words, at these pressures, we approach liquid densities without cryogenics.    Since we do not quite expect to reach liquid densities, the detector needs to expand in scale,  but much of this expansion can be along the ``cheap'' axis of longer drift distances, not the ``expensive'' axis of more or longer anode wires.   

High-pressure TPCs have been explored for underground physics topics including neutrinoless double beta decay~\cite{NEXT}, dark matter~\cite{gehman}, low-energy neutrinos~\cite{aune}, and near detectors for long-baseline neutrino experiments~\cite{nustorm,stainer}.    The benefit of a high-pressure gas is, in general, the freedom to choose the target density.   Low-energy particles have better resolved identities and ranges if the target fluid is of lower density.    In unpressurized labs, only low-total-mass targets have ever been considered, since the experiment must be enclosed in a pressure vessel.   The 8~m$^3$, 20 bar, 280 kg argon target discussed for LNBO is unusually large by these standards.   In a salt cavern, there is no site-related barrier to 10--100kT gas targets, an utter impossibility anywhere else.   We have considerable design freedom: at the time of site/depth-selection, we have freedom to choose a gas pressure over a very wide range (say 20--200 bar).   At a given depth, cavern cooling/heating gives us fairly sensitive control over the gas density during operations: for example, at 100 bar, a 60$^{\circ}$~C temperature swing gives a factor of $\sim$3 change in the density of argon.   We have the freedom to choose a favorable gas mixture, including quenching, wavelength-shifting, Penning, or negative-ion drift components if desired, unlike a cryogenic TPC where non-noble gases freeze out.    On the other hand, it remains to be seen whether long electron lifetimes can be obtained in a warm detector.   The author could not find evidence in the literature for supercritical operation of a gas ionization detector of any kind.

A serious complication comes from the \emph{compressibility} of the target material; we have to consider how to get our instrumentation through the transition from STP to high density without crushing.    This is tricky, and the possible solutions are closely entangled with the solution-mined cavern constraints, so we will present them in some detail.

We suggest four construction approaches.   First, we can deliver TPC subunits underground supported by incompressible fluid for transport, then drain and refill them after emplacement in the cavern.   Such units might be delivered in a fully-built or a partly collapsed state.  Second, we consider (and reject) a system for delivering argon to TPC units during their descent.  Third, we consider filling the entire cavern with the detector gas. Finally, we contemplate dropping our earlier insistence on a perpetually-pressurized cavern, and performing the installation in an air-filled cavern at ambient pressure.     

\subsubsection{Blowout safety}  The addition of pressurized gas to the cavern raises a blowout safety issue.   While gas-pressurized caverns are familiar in the energy industry, these caverns have only small (10--18'') access wells, built entirely as a high-pressure system and permanently capped by high-pressure piping.    We are instead discussing an effectively open shaft which has no pressure-holding capacity other than its water column.    Consider the fate of the argon ``bubble'' released by catastrophic failure of any gas-filled component.   This bubble could rise into the shaft and begin climbing under its own buoyancy.   As it rises, it depressurizes and expands, driving water out of the shaft; a 10-ton gas bubble (expanding to 10,000m$^3$ at STP) will forcefully eject a large fraction of the shaft water.  The loss of water incidentally depressurizes the entire cavern and likely causes failure of all gas-filled vessels there.   To prevent bubbles from rising into the shaft, at the very least we should have a bulkhead door at the cavern collar, perhaps backed by removable hydraulic or thermoplastic plugs.   (Note the impact of such a door on pipe and cable routing.)  For additional blowout safety, the cavern collar should \emph{not} form a ``funnel'' from the cavern into the shaft, but rather have its door at least several meters below the high point of the cavern.    With this, most in-cavern gas releases would deliver an argon bubble to the highest point in the cavern (higher than the lip of the collar) where it can be left alone or vented cautiously via a smaller pipe.     As an extreme anti-blowout measure, we might choose to entomb the detector permanently---after installation, we deposit a cache of appropriate ROVs and spare parts in the cavern, then plug the installation shaft with cement or mud, as is done for abandoned caverns.  The detector would continue to operate but could no longer be easily accessed for repairs or upgrades.    (This may be missing the point if the blowout-risk is most acute during installation, which is the case for some of the plans below.)   

\subsubsection{Fully-assembled, enclosed TPCs}\label{section_captain}

Consider a TPC modeled after CAPTAIN~\cite{captain}.   In CAPTAIN, an octagonal anode assembly is attached to a drift cage, forming a long octagonal or cylindrical drift volume with drift along the axis of symmetry.   Unlike LBNE's rectangular prisms, the CAPTAIN octagons are well-suited to underground assembly because the radius is reasonably well matched to the shape and size of the access shaft. 

For concreteness, we'll discuss a 2m diameter double-sided anode assembly coupled to two 5~m long drift volumes~\footnote{We arrive at 5~m by guessing a worse-than-LBNE argon purity of 400~ppt O$_2$, gas density of 350~kg/m$^3$, and 5~cm/$\mu$s drift.}   The assembly is housed in a 2.2m x 11m steel tube capable of holding 1 bar internal overpressure.    When filled at 350 kg/m$^3$, this will have 9 tonnes of active mass, so we will require 90 such units per kiloton.    A 50 kT detector (4500 units) can be assembled in approximately one year at a pace of 2 hours per unit.    How do we get these TPCs into the cavern?

\paragraph{Incompressible buffers}  When we lower the TPC into the shaft water,  we fill it with an alcohol, fluorocarbon, oil, or another clean and reasonably-incompressible fluid.\footnote{We note that ``incompressible'' is a relative term; most fluids will compress by 1-2\% at 100 bar, which we handle with a  compensation bellows.}   This prevents the housing from crushing during the pressure increase.    We install the TPC in the cavern and attach both top and bottom fill hoses.   Once the TPC is anchored\footnote{After the buffer/argon replacement, the TPC and its ballast will exert a 30+ t buoyancy force on the anchor.} we drain the buffer via the lower hose and replace it with argon via the upper hose and a regulator.   The same two-hose plumbing system then purges the argon repeatedly, sending it to purification plant on the surface.   The buffer fluid needs to be chosen for compatibility with a purification/purge process.  We note that all of these operations involve fairly low pressure differentials across in-cavern vessels, regulators and manifolds; therefore, catastrophic failure of one TPC should not induce the sudden cascading-failure problem that occurs in submerged PMTs.  

\paragraph{Fill-during-descent} Consider a TPC unit which is lowered into the cavern while attached to a long semiflexible argon hose.  The hose is connected to the TPC via a regulator, which matches the TPC pressure to the ambient water pressure.   The shaft descent speed is slow enough to permit the TPC to fill with argon, right up to the final operating pressure, during the entire trip to its installation depth.    This requires substantial additional ballast, since the TPC will begin the trip with ~40 tonnes-force of buoyancy.   The descent speed limitation may be prohibitive.    Moreover, this method presents the worst possible blowout risk: if a TPC fails during descent, it's already in an unprotected open shaft where the argon bubble will cause a blowout, including violent depressurization of the cavern bulkhead and presumably the entire cavern.   Fully-argon-filled TPCs should only be permitted in a pressure-tolerant and capped-off shaft, or one equipped with an active blowout preventer.  As far as the author can determine, blowout prevention is unprecedented for a large shaft.   Although this may merit some further study, skepticism is warranted.   
 
Both of these CAPTAIN-like designs make a physics compromise: they divide the active volume into vertical cylinders, separated by uninstrumented water-filled dead volumes amounting to 20\% of the target mass.  Is this acceptable for a long-baseline neutrino experiment?   Clean detection of electron-neutrino appearance signatures requires, e.g., high efficiency detection of neutral pions, so the answer may be ``no'' although this requires study.    Can we pack the cavern more densely with instrumentation, or with larger contiguously-instrumented volumes?    Rather than attempting to sketch a densely-packable (square or hexagonal) version of the previous scheme, we will sketch a method for filling the whole cavern with the detector gas.

\subsubsection{Dry caverns} 

In these schemes the \emph{entire} lined cavern, not an individual series of TPC units, is an argon vessel.  This imposes cleanliness and materials constraints on the liner and all of the installed equipment (cables, hoses, electronics) which previously were exposed only to the cavern fluid.   In terms of blowout prevention, the cavern/shaft bulkhead door and its plugs have become a ``single point'' safety-critical item.    With proper argon injection and water withdrawal, the bulkhead door will not experience large pressure forces, but we must ensure that this remains true at all times, like during a downtime of the argon-supply system or failure of a cavern cooling system.   

\paragraph{Cavern wet for installation} Consider a lined cavern, flooded with clean water, into which TPC components (including LNBE-style rectangles, folded as discussed below) are delivered \emph{without} a pressure enclosure.    All of the gas-facing components (wires, drift cages, insulators, etc.) are simply immersed in the water during descent, allowing us to use an ROV for installation.   Later, argon will replace the water in the entire cavern at once.    In previous cases, we always installed a neutrally-buoyant detector assembly and then \emph{increased} its buoyancy during the final step, which motivated the previous suggestions to have detectors tied down to the cavern floor.   In this case, a detector assembly which is neutrally-buoyant in water will be negatively-buoyant in argon.    The detectors should therefore either be (a) suspended from the roof or (b) rigidly affixed (not cabled) to the floor so that nothing shifts position during the argon fill.

\begin{figure*}
\begin{center}  \includegraphics[width=0.8\linewidth]{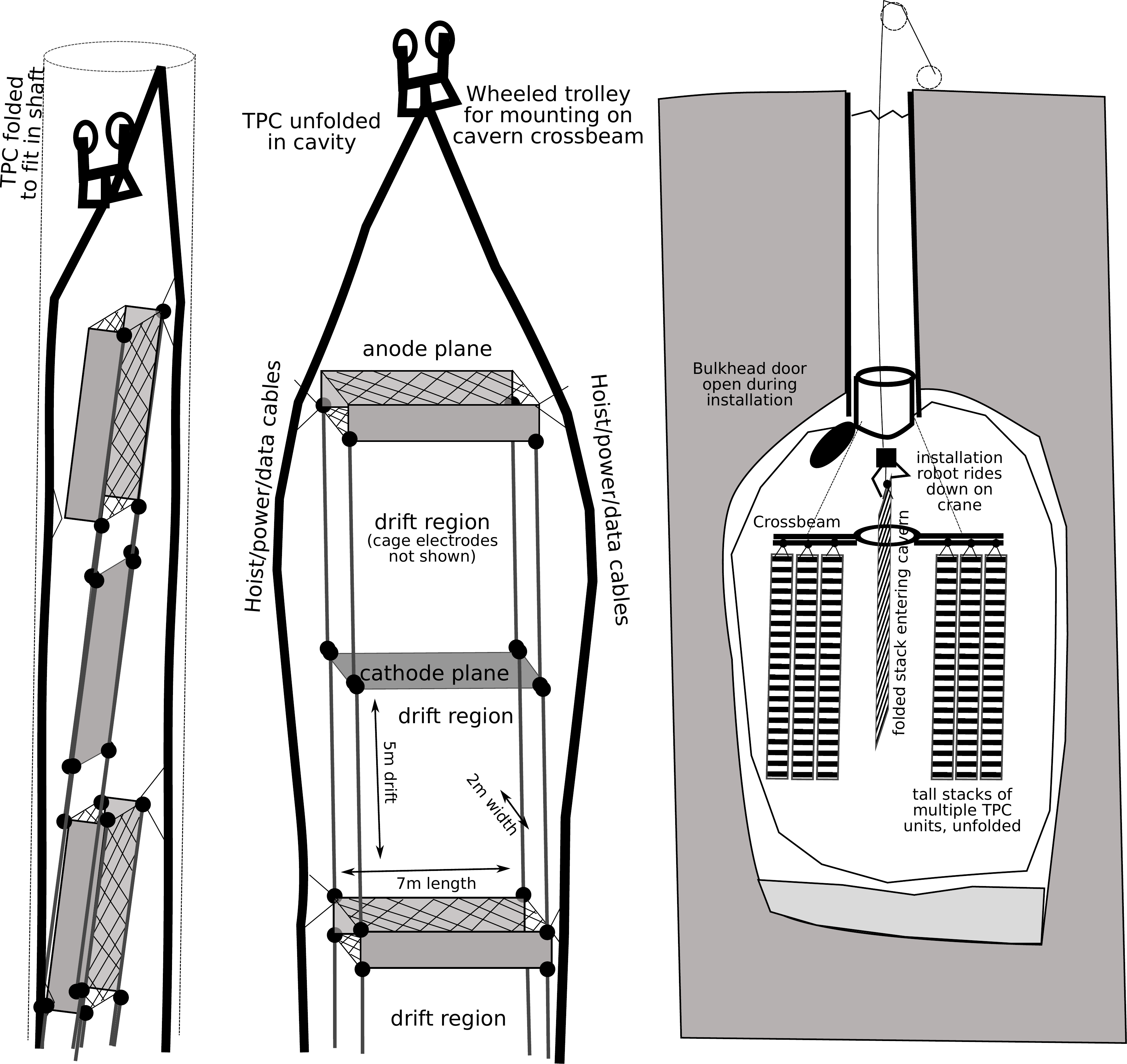}\end{center}
\caption{Folding, unfolding, and installation of an LBNE-like TPC.  Left: a tall stack of TPC anodes and cathodes ``folds'' via a hinge that collapses the drift cage.   Right: A long I-beam, spanning the cavern, is hung from a mounting structure on the collar.  Each TPC stack includes a wheeled trolley which can slip onto the beam.  Each stack is lowered into the center of the cavern by a crane which includes two hydraulic manipulator arms.   Using the manipulators, the trolley is pushed onto a crossbeam, detached from the crane, and unfolded.   A more-complicated, multi-piece crossbeam would allow a fuller 3-D space-filling arrangement of units.}\label{argon_schematic}
\end{figure*}

Since this process doesn't require a pressure vessel, gas balloon, etc., it offers a reasonably easy route towards ``folding'' an LBNE-sized TPC subassembly for delivery down a shaft.   Specifically, the drift cage can be hinged to collapse the drift region and drift cage, perhaps as shown in figure \ref{argon_schematic}.

Once the detector assemblies are all emplaced, the cavern bulkhead door is closed and the cavern is drained of water and filled with argon.   The drain procedure is conducted from the top down, with high-pressure argon injected on top, and driving the water into a pipe that reaches to the bottom of the cavern.    Extensive argon flushing/recycling, and perhaps even a moderate-temperature ``bakeout'' of the cavern, will be needed to achieve the desired gas purity.   If water turns out to be unsuitable, another fluid must be used.   The enclosed-TPC buffer-fluid proposal required the surface lab to stock only a \emph{few tons} of the buffer fluid, which was recycled over and over as new units were filled, emplaced, and drained.   Here, the user must fill the \emph{entire cavern} temporarily with the chosen buffer fluid, which may be costly.

\paragraph{Dry installation}
Finally, it appears to be the case that an appropriately-designed cavern can be safely depressurized, left unsupported for a short period, then repressurized.   Can we install a TPC in an open, dry, atmospheric-pressure cavern?    It would have obvious advantages for argon handling and purity, especially if we install enclosed CAPTAIN-like units, as in section~\ref{section_captain}, which can be cleaned under vacuum on the surface.    However, the robotic/automated installation now requires particularly precise and reliable overhead-crane choreography, because we no longer have access to a swimming ROV or to buoyancy support for components.   The procedure suggested in figure~\ref{argon_schematic} appears compatible with this idea.

\subsection{Small detectors: dark matter, double-beta decay, solar neutrinos, etc.}
We have outlined several ways to construct an detector in a cavern piece by piece.    Some small experiments might have a ``core'' technology that can be inserted intact down a shaft.  For example,  LZ~\cite{lz} is designed as a 1.5~m diameter cylindrical TPC carrying 7~t of liquid xenon.   At the Sanford underground lab it will be installed in an 8~m diameter tank carrying water shielding and liquid-scintillator veto systems.    Very briefly, we suggest a salt-cavern option for this category of experiment: a water shield and veto system can be built using the techniques outlined previously, and the core apparatus is lowered in one piece into the veto.  Are there any advantages to this idea?   Are there any technical barriers?  

Advantages include the following: 
\begin{itemize}
\item Low-cost green-field development, with infrastructure costs that scale appropriately with the size of the experiment.  We have already mentioned the desire for a Southern Hemisphere site for followup of any dark matter annual modulation result.   
\item Extremely large water shielding and veto volumes---tens of meters rather than a few meters. 
\item It is possible to develop salt caverns at great depths where there is presently fairly little (2~km) or no (3~km) mined lab space at present.  
\item Experiments requiring high pressures (bubble chambers~\cite{COUPP} with a specific temperature/pressure requirements, and high-pressure TPCs~\cite{NEXT,gehman,aune}) might view solution-mined caverns solely as giant pressure vessels.
\end{itemize}

Technical challenges include the following:   
\begin{itemize}
\item  It is not possible to build a lightweight vacuum cryostat that withstands high pressures.  Millikelvin-cooled detectors like CDMS are therefore out of the question.   Noble liquids may be possible on a small scale; the MiniCLEAN collaboration investigated~\cite{nikkel} the possibility of cryostat-less detector insulated by water ice.    
\item PMTs in pressure housings are more radioactive than bare PMTs.   
\item Detector support systems need to be located either in the cavern or a full shaft-length away; neither is convenient.
\end{itemize}

\section{Translating salt-cavern technologies to the seafloor}\label{ocean}

Once we are designing detectors capable of withstanding high pressure, the question arises: are they suitable for the ocean floor, too?   In terms of pressure protection the answer is clearly yes.  Since we are already engaged in speculation, we will discuss some of the undersea applications of the methods we propose for salt cavities.    The ability to operate on the sea floor gives experimenters the ultimate freedom of site-selection and experiment-size.   There are no geotechnical limits on the size, shape, or volume of a detector---only economic and mechanical-engineering limits.   Mobile detectors are particularly interesting.    Seafloor detectors can be positioned any desired distance from an accelerator, and perhaps even repositioned from on-axis to off-axis sites in the same beam.   A reactor antineutrino detector can move towards and away from a reactor to measure an oscillation profile.    A geoneutrino experiment can do a ``transect'' of the ocean to survey the Earth's mantle~\cite{hanohano}.    

The engineering constraints of ocean deployment are different than those of a solution-mined cavern.   In a cavern, we have a difficult and highly-constrained entryway to the detector via a shaft, but we have a permanent and stable presence at the top of this shaft.   It is clearly possible to view detector construction as a multi-year process including a large number of steps, arbitrarily-slow steps like the careful inflation of a liner, etc., all with more-or-less uninterrupted access to fluids, gases, vacuum, power, cranes, loading docks, etc..   For an ocean-floor experiment, there is no shaft-geometry constraint but rather a time constraint.   One chooses to assemble an easy-to-deploy detector on shore, then wait for a window of good weather to run a simple and interruption-tolerant deployment cruise.  

The inflatable hodoscope described in section~\ref{ls} is fairly straightforward to adapt to the seafloor.   The detector's ``deflated'' state gives it a mass compatible with standard marine construction cranes, but since the deflated shape is less constrainted there is no barrier to a 90$^\circ$ stereo angle.  Scintillator can be delivered to the installed detector via a buoyed riser pipe.  

The pressurized argon TPC is an intriguing option for seafloor deployment.   There is no longer a blowout hazard: a subdetector rupture at sea is a waste of argon but not a general disaster.    One can go deeper into the ocean than into a stable salt cavern, and find consistently low ambient temperatures, leading to quite high gas densities (750 kg/m$^3$ at 4$^\circ$ and 400 bar, for example).   With no shaft-size constraint, one can contemplate a Hanohano-like deployment where a very large detector is constructed in one piece, filled with an incompressible buffer fluid (and made neutrally buoyant) for descent and anchoring, and later flushed with gas, either via a riser pipe or possibly a pipeline to shore.

Interestingly, a hydrogen-filled or methane-filled TPC could reach a hydrogen density (28~mol~H/l or 68~mol~H/l, respectively, at 400 bar) only somewhat lower than that of liquid scintillator (110~mol~H/l), and therefore might make an excellent geoneutrino detector.

\section{Conclusions}

We have introduced the idea of salt-cavern sites for large underground detectors.   These sites impose some noteworthy construction and operations constraints, but if the constraints can be met they may serve as extremely low-cost facilities for long-baseline neutrino physics, supernova neutrinos, proton decay searches, and other underground science.    We sketch out some specific, albeit speculative, ideas for salt-cavern installation of water Cerenkov detectors, scintillating hodoscopes, and high-pressure TPCs, without uncovering any  obviously-insurmountable barriers.   An ambitious program of detector design, engineering, and pilot experiments is called for, with input from physicists, engineers, and cavern-storage industry professionals, if we wish to construct large salt cavern experiments in the future.

\section{Acknowledgements}
The author is indebted to  Sarah Bagby, Stephen Bauer, Kerry DeVries, Leon Mualem, and Ryan Patterson for helpful discussions and to Jocelyn Monroe, James Nikkel, Tom Stainer, and  Chris Walter  for pointers to references.

\bibliographystyle{JHEP_bm}
\bibliography{saltmine}

\end{document}